\documentclass [12pt] {report}
\usepackage{amssymb}
\usepackage[dvips]{graphicx}
\newcommand {\eqdef} {\stackrel{\rm def}{=}}

\newcommand {\sign} {\mathop{\rm sign}\nolimits}

\newcommand {\arccosh} {\mathop{\rm arccosh}\nolimits}
\newcommand {\arctanh} {\mathop{\rm arctanh}\nolimits}

\newcommand {\D}[2] {\displaystyle\frac{\partial{#1}}{\partial{#2}}}

\newcommand {\al} {\alpha}

\newcommand {\si} {\sigma}
\newcommand {\Si} {\Sigma}

\newcommand {\de} {\delta}
\newcommand {\De} {\Delta}

\newcommand {\iy} {\infty}
\newcommand {\prtl} {\partial}
\newcommand {\fr} {\displaystyle\frac}

\newcommand {\be} {\begin{equation}}
\newcommand {\ee} {\end{equation}}
\newcommand {\ba} {\begin{array}}
\newcommand {\ea} {\end{array}}
\newcommand {\bp} {\begin{picture}}
\newcommand {\ep} {\end{picture}}
\newcommand {\bc} {\begin{center}}
\newcommand {\ec} {\end{center}}

\newcommand {\bt} {\begin{tabular}}
\newcommand {\et} {\end{tabular}}
\newcommand {\lf} {\left}
\newcommand {\rg} {\right}

\newcommand {\nin}{\noindent}

\newcommand {\cE} {{\cal E}}
\newcommand {\cF} {{\cal F}}

\newcommand {\cI} {{\cal I}}

\newcommand {\cK} {{\cal K}}

\newcommand {\cQ} {{\cal Q}}
\newcommand {\cR} {{\cal R}}
\newcommand {\cS} {{\cal S}}
\newcommand {\cV} {{\cal V}}

\newcommand {\bP} {{\bf P}}

\newcommand {\bR} {{\bf R}}

\newcommand{\mP}{\mbox{$|{\bf P}|$}}

\newcommand{\mR}{\mbox{$|{\bf R}|$}}
\newcommand {\ses} {\medskip}
\newcommand {\pgbrk}{\pagebreak}

\def\2#1#2#3{{#1}_{#2}\hspace{0pt}^{#3}}
\def\3#1#2#3#4{{#1}_{#2}\hspace{0pt}^{#3}\hspace{0pt}_{#4}}
\newcounter{sctn}
\def\sec#1.#2\par{\setcounter{sctn}{#1}\setcounter{equation}{0}
                  \noindent{\bf\boldmath#1.#2}\bigskip\par}
\begin {document}

\begin {titlepage}

\vspace{0.1in}

\begin{center}
{\large \bf Finsleroid--Relativistic Space  Endowed With Scalar Product
}\\
\end{center}

\vspace{0.3in}

\begin{center}

\vspace{.15in}
{\large G. S. Asanov\\}
\vspace{.25in}
{\it Division of Theoretical Physics, Moscow State University\\
119992 Moscow, Russia\\
(e-mail: asanov@newmail.ru)}
\vspace{.05in}

\end{center}

\begin{abstract}

\ses

 When a single time-like vector is distinguished geometrically
to present the only preferred direction in
  extending the  pseudoeuclidean geometry,
the hyperboloid may not be regarded as an exact carrier of the
unit-vector image. So under respective conditions one may expect
that some time-assymetric figure should be substituted
with the  hyperboloid. To this end we shall use the {\it pseudo-Finsleroid}.
The spatial-rotational invariance (the $\cal P$-parity) is retained.
The constant negative curvature is the fundamental property of the
 pseudo-Finsleroid surface.
The present paper develops the approach in the direction of evidencing the concepts of angle,
scalar product, and geodesics.
In Appendices we shortly outline the basic aspects that stem from the
choice of the Finsleroid-relativistic metric functions.

\end{abstract}

\end{titlepage}

\vskip 1cm

\nin
{\bf  1. Introduction}

\bigskip

The  attempts  to introduce the concept of angle in the
Minkowski or Finsler spaces [1-5] were steadily encountered with
difficulties.
In the present paper we follow and realize the idea
that  the angle
should be obtainable from the geodesics through postulating the Cosine Theorem
of the standard form.

In the sequel the abbreviations FMF, FMT, and FHF will be used for the
Finsleroid metric function,  the associated Finsleroid metric
tensor, and the associated Finsleroid Hamiltonian function,
respectively. The notation  $\cE_g^{SR}$ will be applied to
the Finsleroid-relativistic  space
with the subscripts $``SR"$ meaning ~``special-relativistic".~
The characteristic parameter $g$,
which measures the deviation of the $\cE_g^{SR}$-geometry from its
pseudoeuclidean precursor,
 may take on the values over the range
$(-\infty,+\infty)$;\,
at $g=0$  the space is reduced to become the  ordinary pseudoeuclidean one.

Section 2 is devoted to
presenting the key and basic concepts determined by geodesics and  angle.
The equations (A.30)-(A.31) for the
$\cE^{SR}_g$-geodesics  prove to admit a simple and
explicit general solution (the convenient  method of solution is to follow closely
  the
 method used in the paper [6] in the positive-definite case),
from  which  the angle (Eq. (2.1)) can be obtained.
The respective scalar product (Eq. (2.2)) ensues.
The solution
with fixed points, as well as the initial-date solution, are both presented explicitly.
An essential non-pseudoeuclidean  feature is that the  $\cE^{SR}_g$-geodesic curves
 are not flat in general.

Appendix A gives an account of the
notation and conventions for the space
 $\cE^{SR}_g$
and introduces the initial concepts and definitions that are
required. The space is constructed by assuming an axial symmetry
and, therefore, incorporates a single preferred timelike direction, which
we shall often refer as the $T$-axis (or the $R^0$-axis). After preliminary
introducing a characteristic quadratic form $B$, which is distinct
from the pseudoeuclidean  quadratic form by presence of a mixed term (see
Eq. (A.10)), we define the FMF $F$ for the space $\cE^{SR}_g$ by
the help of the formulae (A.12)-(A.13).
Next, we present the results of calculating the  basic tensor quantities of the space.
As well as in the pseudoeuclidean geometry  the locus of the unit
vectors issuing from fixed  point of origin is the unit hyperboloid, in
the $\cE_g^{SR}$-geometry under development the locus is the
boundary (surface) of the Finsleroid. We call the boundary the
Indicatrix. It can rigorously be proved that the
 Indicatrix  is  regular  and locally  convex.
 The value of the curvature depends on the parameter
$g$ according to the simple law (A.29). The determinant of the
associated FMT is strongly negative in accordance with Eqs.
(A.18)-(A.19).
The consideration can conveniently be
converted  into the co-approach. The explicit form of
 the associated  FHF
 is entirely similar to the form of the FMF $F$ up to the
substitution of $-g$ with $g$.
The $\cE^{SR}_g$-space has an auxiliary quasi-pseudoeuclidean structure,
 which is deeply inherent in the development.
Appendix B introduces for  the $\cE^{SR}_g$-space the
quasi-pseudoeuclidean map under which the pseudo-Finsleroid goes into the unit
hyperboloid. The quasi-pseudoeuclidean space is simple in many aspects,
 so that relevant transformations make reduce various calculations.


\pgbrk

\setcounter{sctn}{2}
 \setcounter{equation}{0}
 {\bf 2.  Scalar product, angle and geodesics}

 \ses \ses

 Given two four-dimensional vectors $R_1\in \cV_g$ and $R_2\in
\cV_g$.
 Let us define the {\it
 $\cE_g^{SR}$--scalar product}
 \be
<R_1,R_2>~:=F(g;R_1)F(g;R_2)
\cosh
\Bigl[ \fr1h\arccosh\fr {
A(g;R_1)A(g;R_2)-h^2r_{be}R_1^bR_2^e } {
\sqrt{B(g;R_1)}\,\sqrt{B(g;R_2)} } \Bigl]
 \ee
so that the $\cE_g^{SR}$--{\it angle}
 \be
 \al(R_1,R_2)~: = \fr1h\arccosh \fr
{ A(g;R_1)A(g;R_2)-h^2r_{be}R_1^bR_2^e } {
\sqrt{B(g;R_1)}\,\sqrt{B(g;R_2)} }
 \ee
is appeared between the vectors $R_1$
and $R_2$; the functions $B, K$, as well as  $A$ can
be found in Appendix A.

The general solution
\be
 R^p=R^p(s)
\ee
to the  $\cE_g^{SR}$-space geodesic equations
(presented by
 Eqs. (A.30)-(A.31)) proves to be given explicitly by means of the components
 \be
  R^0(s)=(t^0(s)+\fr12Gm(s))/k(s),
\qquad R^a(s)=\fr1ht^a(s)/k(s)
 \ee
with
 \be
t^0(s)=\fr{F_s}{\sinh(h\al)}\Bigl[\fr{A(g;R_1)}{\sqrt{B(g;R_1)}}\sinh(h(\al-\nu))
+ \fr{A(g;R_2)}{\sqrt{B(g;R_2)}}\sinh(h\nu)\Bigr],
 \ee
\ses
 \be
t^a(s)=h\fr{F_s}{\sinh(h\al)}\Bigl[\fr{R_1^a}{\sqrt{B(g;R_1)}}\sinh(h(\al-\nu))
+ \fr{R_2^a}{\sqrt{B(g;R_2)}}\sinh(h\nu)\Bigr], \ee
\ses
where
  \be
F_s=\sqrt{(F(g;R_1))^2+2bs +s^2},
\ee
\ses
 \be b=F(g;R_1)
\sqrt{1+\lf(\fr{F(g;R_2)\sinh\al}{\De s }\rg)^2},
 \ee
and
 \be
k(s)=\Bigl|\fr{t^0(s)-m(s)}{t^0(s)+m(s)}\Bigr|^{-G/4}, \qquad
m(s)=\sqrt{r_{ab}t^a(s)t^b(s)}.
 \ee
The intermediate angle $\nu$ is equal to
 \be \nu=\arctanh\fr{sF(g;R_2)\sinh\al}{F(g;R_1)\De
s+[F(g;R_2)\cosh\al-F(g;R_1)]s} \ee
and is showing the property
$$
\nu_{{ |}_{s=0}}=\al.
$$
Along the geodesics,
we have
 \be
F(g;R(s))=F_s,
\ee
so that  the behaviour law for the squared FMF $F^2$ is
quadratic with respect to the parameter $s$:
\be
(F(g;R(s)))^2=a^2+2bs+s^2\equiv (b+s)^2-(b^2-a^2);
 \ee
 $a$ and $b$ are two constants of integrations
 with $a=F(g;R_1)$.
 It is assumed that
\be
b^2-a^2\ge 0.
\ee

\pgbrk

The picture symbolizes the role which the angles (2.2) and
(2.10) are playing  in featuring the geodesic line $C$ which joins two points
$P_1$ and $P_2$.

\ses \ses

\begin{figure}[h]
 \centering
\includegraphics[width=9cm]{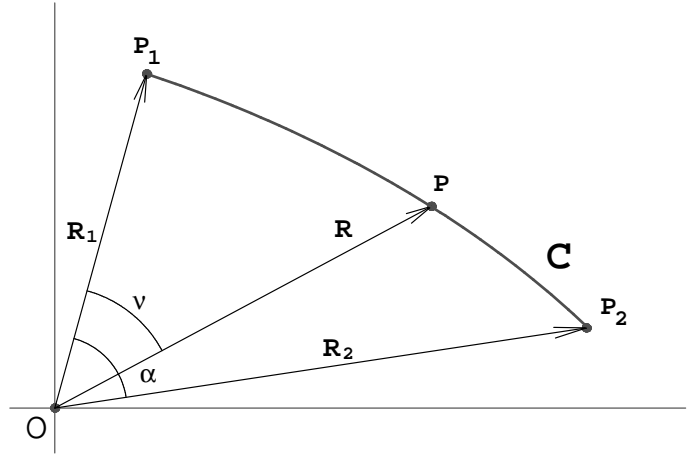}
\caption{\small The geodesic $C$ and the angles $\al=\angle P_1OP_2$ and
$\nu=\angle P_1OP$}
\end{figure}

\ses \ses


On this way  the following substantive items can be arrived at.

\ses \ses

\nin {\it The $\cE_g^{SR}$-Case Two-Point Distance $\De s$}:
\ses\\
\be (\Delta s)^2 = (F(g;R_1))^2 + (F(g;R_2))^2 - 2
F(g;R_1)F(g;R_2) \cosh \al.
 \ee

\ses \ses

\nin {\it The $\cE_g^{SR}$-Case Scalar Product }
\ses\\
\be
 <R_1,R_2>=F(g;R_1)F(g;R_2) \cosh \al. \ee

\ses

\nin {\it At equal vectors, the reduction}
\be
 <R,R>= (F(g;R))^2
\ee
 takes place, that is, the two-vector scalar product (2.1)
reduces  exactly  to the squared FMF.

\ses \ses

\nin {\it The $\cE_g^{SR}$-Case Orthogonality }
\ses\\
\be <R,R^{\perp}>=0. \ee

\ses \ses


Under the identification
 \be
 | R_1 \ominus R_2 |
 =\De s
 \ee
the formula (2.12) can be read as

\ses
\ses
\nin {\it The $\cE_g^{SR}$-Case Cosine Theorem }
\ses\\
\be | R_1 \ominus R_2 |^2 = (F(g;R_1))^2 + (F(g;R_2))^2 - 2
<R_1,R_2>. \ee

\ses \ses

\clearpage

From this we can also conclude that

\ses\ses

\nin {\it The $\cE_g^{SR}$-Case Pythagoras Theorem }
\ses\\
\be | R \ominus R^{\perp} |^2 = (F(g;R))^2 + (F(g;R^{\perp}))^2
\ee
 holds
fine.

\ses\ses

The symmetry
 \be | R_1 \ominus R_2 | = | R_2
\ominus R_1 | \ee
is obvious.


\ses\ses

NOTE.
One can easily execute the formula (2.14) from the representation (2.7)
if one inserts (2.8) in (2.7), takes the case $s=\De s$,  uses the equality
$$F(g;R_2)=F_{\De s}$$
(see (2.11)), and resolves the resultant equation to find ($\De s)^2$.

\ses\ses

Particularly, from (2.2) it directly ensues that  the value of the
angle  $\al$ formed by  a vector $R$ with the pseudo-Finsleroid
$R^N$--axis is given by
 \be
  \al = \fr1h\arccosh \fr { A(g;R) } {
\sqrt{B(g;R)}\ }, \ee
where $A$ is the function defined by Eq. (A.32), and with $(N-1)$--dimensional equatorial
$\{{\bf R}\}$--plane of the pseudo-Finsleroid is prescribed as
 \be
  \al =
\fr1h\arccosh \fr { L(g;R) } { \sqrt{B(g;R)}\ },
 \ee
where $ L$ is
the function given by Eq. (A.33).


\ses \ses

\begin{figure}[h]
\centering \hbox to \textwidth{\hfill
\includegraphics[height=4cm,width=6.5cm]{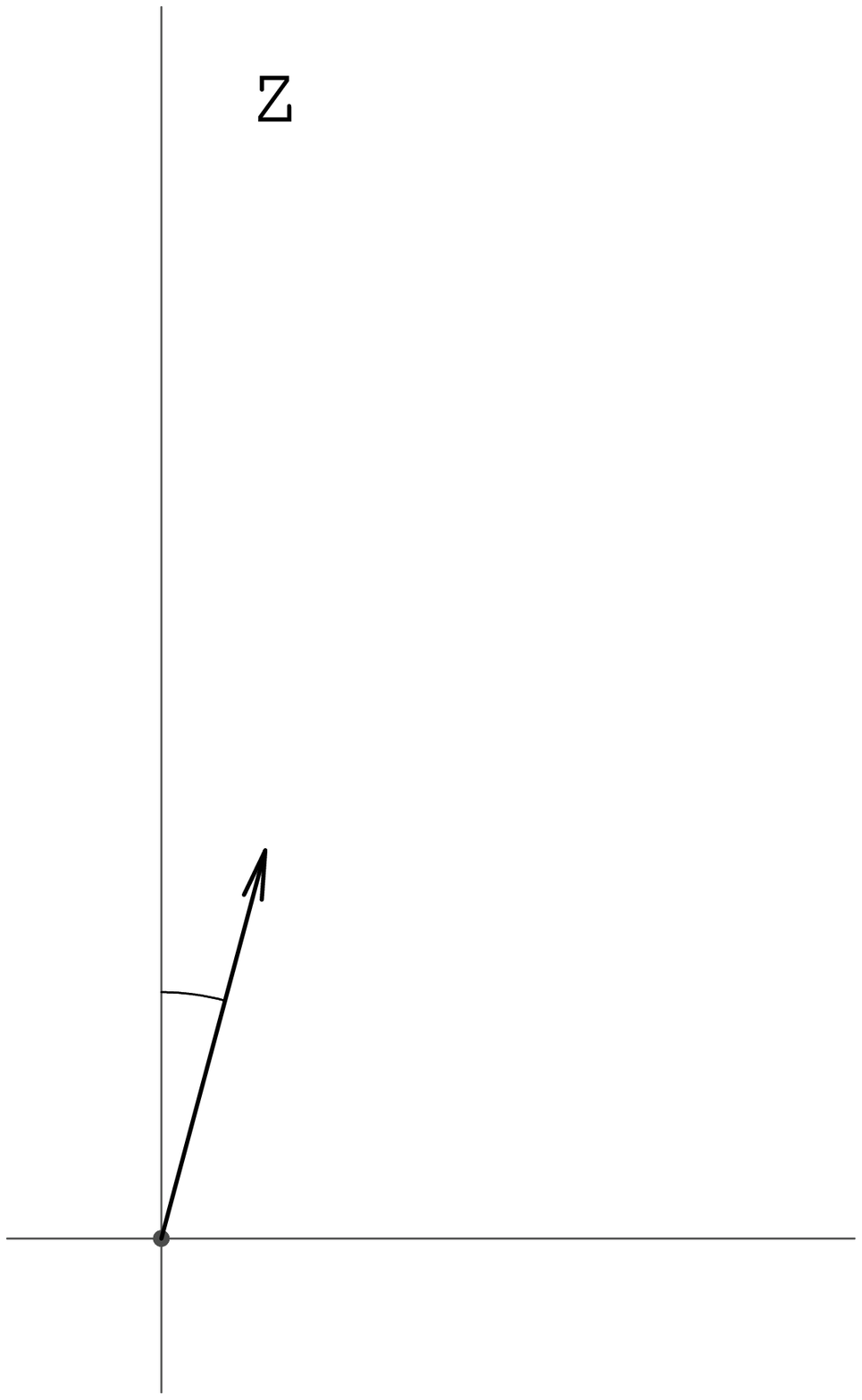}
\hfill
\includegraphics[height=4cm,width=6.5cm]{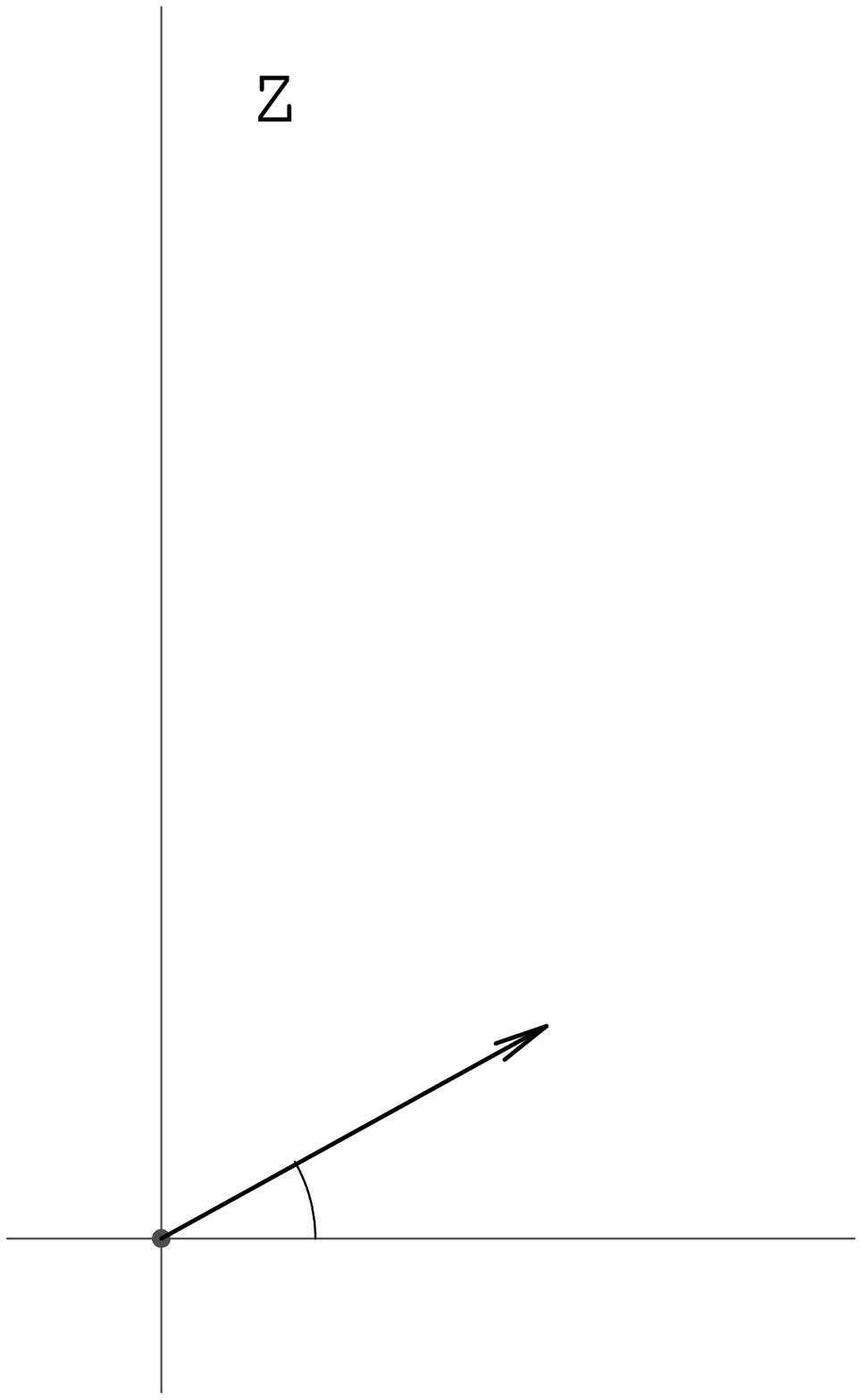}
\hfill}

\caption{\small The angle cases (2.22) and (2.23), respectively.}
\end{figure}

\ses\ses

\clearpage

Just the similar relationships are obtained in the co--approach.
Given two co-vectors $P_1\in \hat\cV_g$ and $P_2\in
\hat\cV_g$,
the analogue of the formula (2.1) is the {\it $\hat\cE_g^{SR}$-scalar product}
 \be
<P_1,P_2>=H(g;P_1)H(g;P_2) \cosh \Bigl[ \fr1h\arccosh\fr {
\hat A(g;P_1)\hat A(g;P_2)-h^2r^{be}P_{1b}P_{2e}} {
\sqrt{\hat B(g;P_1)}\,\sqrt{\hat B(g;P_2)} } \Bigl],
 \ee
 \ses
which corresponds to the {\it $\hat\cE_g^{SR}$-angle}
 \be
 \hat\al(P_1,P_2) = \fr1h\arccosh \fr
{\hat A(g;P_1)\hat A(g;P_2)-h^2r^{be}P_{1b}P_{2e} } {
\sqrt{\hat B(g;P_1)}\,\sqrt{\hat B(g;P_2)} }.
 \ee
The complete analogue
 \be
 |P_1 \ominus P_2 |^2 =
(H(g;P_1))^2 + (H(g;P_2))^2 - 2H(g;P_1)H(g;P_2) \cosh \hat\al
 \ee
to the formula (2.13) and the symmetry
\be
 | P_1 \ominus P_2 | = |P_2 \ominus
P_1 |
 \ee
hold fine.

\ses

 In the pseudoeuclidean limit proper, the right-had parts in
 (2.2) and (2.25)
takes on the ordinary pseudoeuclidean form:
$$
 \al(R_1,R_2)_{{\Bigl |\Bigr.}_{g=0}} = \arccosh \fr
{ R_1^0R_2^0-r_{be}R_1^bR_2^e } {
\sqrt{(R^0_1)^2-r_{be}R_1^bR_1^e}\,\sqrt{(R_2^0)^2-r_{be}R_2^bR_2^e} }
$$
\ses
и
$$
 \hat \al(P_1,P_2)_{{\Bigl |\Bigr.}_{g=0}} =
 \arccosh
 \fr
{
 P_{10}P_{20}-r^{be}P_{1b}P_{2e}
  }
{
\sqrt{(P_{10})^2-r^{be}P_{1b}P_{1e}}\,
\sqrt{(P_{20})^2-
r^{be}P_{2b}P_{2e}}
}.
$$

Using (2.5) and (2.6) in (2.4) yields
 \be
  R^p(s)= \fr{F_s}{k(s)\sqrt{B(g;R_1)}} \fr{
\sinh(h(\al-\nu))} { \sinh(h\al) } R^p_1 +
\fr{F_s}{k(s)\sqrt{B(g;R_2)}} \fr{ \sinh(h\nu)} { \sinh(h\al) }
R^p_2 +X(s)\de^p_N
 \ee
with
 \be
 X(s)= -\fr 12g\fr{F_s}{k(s)} \Bigl[
\fr{ \sinh(h(\al-\nu))} { \sinh(h\al) } \fr{q_1} {\sqrt{B(g;R_1)}} +
\fr{ \sinh(h\nu)} { \sinh(h\al) } \fr{q_2} {\sqrt{B(g;R_2)}}
-\fr{m(s)}{hF_s} \Bigr].
 \ee
 Since the additional term
$X(s)\de^p_N$ has appeared
 in the right-hand part of (2.28),
 and the right-hand part in (2.29) does not vanish identically,
we are to conclude that in general the vector $R^p(s)$ is not
spanned by two end vectors $R^p_1$ and $R_2^p$. Therefore,
{\it in general the $\cE_g^{SR}$-geodesic curves obtained are not plane curves}.

\ses\ses

The velocity
components
\be
 U^p(s) \eqdef\fr{dR^p}{ds}
 \ee
can conveniently be deduced from the equalities
 \ses
  \be
   U^p(s)=\mu^p_j(g;{\bf
t}(s))\fr{dt^j}{ds},
 \ee
where
$
\mu^p_j
$
are the functions that are the quasi-pseudoeuclidean functions (B.13).
Calculations show that
 \be
U^p(s)=\fr{b+s}{(F_s)^2}R^p(s) + \fr{hF(g;R_1)F(g;R_2)\sinh\al}{
k_sF_s\sinh(h\al)\,\De s} T^p(s)
\ee
 with
  \be
   T^N(s)=
\fr{A(g;R_2)}{h^2\sqrt{B(g;R_2)}}\cosh(h\nu) -
\fr{A(g;R_1)}{h^2\sqrt{B(g;R_1)}}\cosh(h(\al-\nu))
 \ee
 and
 $$
T^a(s)= \fr{1}{\sqrt{B(g;R_2)}} \lf[
R_2^a+\fr12g\fr{A(g;R_2)}{h^2q}R^a(s) \rg] \cosh(h\nu)
$$
\ses
\be
 -
\fr{1}{\sqrt{B(g;R_1)}} \lf[ R_1^a+\fr12g\fr{A(g;R_1)}{h^2q}R^a(s)
\rg]\cosh(h(\al-\nu)).
 \ee

\ses\ses

It follows that
\be {U^p(s)}_{{\Bigl |\Bigr.}_{g=0}}
=\fr{R_2^p-R_1^p}{\De s},
 \ee
 and the contraction
 \be R_p(s)U^p(s)=b+s
  \ee
is valid, where $R_p$ are the covariant vector components (defined below (A.15)).
 Also,
  \be
g_{pq}(g;R(s))U^p(s)U^q(s)=1.
 \ee

\ses

The {\it initial-data solution}
 \be
 R_2^p=R_2^p(g;R_1,U_1,\De s)
  \ee
can also be explicitly found, namely we get
\be
 R^p_2=\mu^p(g;t_2)
  \ee
  with the functions
 \be
 t_2^i=z(\De
s)\si^i(g;R_1)+n(\De s)\si^i_q(g;R_1)U^q_1,
\ee
\ses
 \be z(\De
s)=\fr1h \fr{\sinh(h\al)}{\sinh\al} + \fr{F(g;R_2)}{F(g;R_1)}
\lf[\cosh(h\al)-\fr1h \fr{\sinh(h\al)}{\sinh\al}\cosh\al\rg],
 \ee
\ses
\be
 n(\De s)=\fr1h \fr{\sinh(h\al)}{\sinh\al} \De s,
\ee
\ses
\be
F(g;R_2)=\sqrt{(F(g;R_1))^2+2b\De s+(\De s)^2},
 \ee
 \ses
\ses
\be b=R_{1q}U^q_1,
 \ee
and the angle value $\al$ can be taken as
\be
 \al=\arccosh \fr{(F(g;R_1))^2+b\De
s}{F(g;R_1)F(g;R_2)}.
\ee
The functions $\si^i$ and $\si^i_q$
are the inverses to $\mu^i$ and $\mu^q_i$, respectively.


\pgbrk



\setcounter{sctn}{1} \setcounter{equation}{0}

{\nin\bf Appendix A. Basic properties of the space $\cE_g^{SR}$}

\ses \ses

Searching for extension of the pseudoeuclidean geometry
in due Finsler-relativistic  way,
we should
adapt constructions to the following decomposition
 \be
\cV_g=\cS_g^+\cup \Si_g^+\cup{\cR_g}\cup\Si_g^-\cup\cS_g^-, \ee
which sectors relate to the cases when the  contravariant vector
 $R\in\cV_g$ is respectively
 future--timelike, future--isotropic, spacelike, past--isotropic, and
 past--timelike.
The respective co-analogue for the covariant vectors (momenta) $P\in\hat\cV_g$
reads \be \hat\cV_g=
\hat\cS_g^+\cup\hat\Si_g^+\cup\hat\cR_g\cup\hat\Si_g^-\cup\hat\cS_g^-.
\ee

With this purpose, we introduce the following convenient notation:
\be G= g/h, \ee
\medskip
\be \qquad h\eqdef\sqrt{1+\fr14g^2}, \ee \ses \be g_+=-\fr12g+h,
\qquad g_-=-\fr12g-h, \ee
\medskip
\be G_+=g_+/h\equiv -\fr12G+1, \qquad G_-=g_-/h\equiv -\fr12G-1,
\ee
\medskip
\be g^+=1/g_+=-g_-,  \qquad  g^-=1/g_-=-g_+, \ee
\medskip
\be g^+=\fr12g+h, \qquad g^-=\fr12g-h, \ee
\medskip
\be G^+=g^+/h\equiv \fr12G+1, \qquad G^-=g^-/h\equiv \fr12G-1. \ee

We shall decompose vectors to select the timelike components and the
three-dimensional spatial components: $ R=\{R^0,\bR\}$ и $ P=\{P_0,\bP\}. $
In terms of the forms \be B(g;R)=-\lf(R^0+g_-|{\bf
R}|\rg)\lf(R^0+g_+|{\bf R}|\rg)
\equiv-\lf((R^0)^2-gR^0\mR-\mR^2\rg), \ee
\medskip
\be \hat B(g;P)=-\lf(P_0-\fr{|{\bf P}|}{g^+}\rg)\lf(P_0-\fr{|{\bf
P}|}{g^-}\rg) \equiv-\lf((P_0)^2+gP_0\mP-\mP^2\rg), \ee
\ses\\
all the sectors entered the decompositions (A.1) and (A.2)
can be embraced by one FMF
 \be
F(g;R)=\sqrt{|B(g;R)|}\,j(g;R)=\lf|R^0+g_-|{\bf
R}|\rg|^{G_+/2}\lf|R^0+g_+|{\bf R}|\rg|^{-G_-/2}, \ee where \be
j(g;R)=\lf|\fr{R^0+g_-|{\bf R}|}{R^0+g_+|{\bf R}|}\rg|^{-G/4}, \ee
\ses\\
and one FHF \be H(g;P)=\sqrt{|\hat
B(g;P)|}\,\hat j(g;P)=\lf|P_0-\fr{|{\bf P}|}{g^+}\rg|^{G^+/2}
\lf|P_0-\fr{|{\bf P}|}{g^-}\rg|^{-G^-/2}, \ee where \be \hat
j(g;P)=\lf|\fr{P_0-\fr{|{\bf P}|}{g^+}}{P_0-\fr{|{\bf
P}|}{g^-}}\rg|^{G/4}. \ee

\ses

By following the methods of the Finsler geometry,
we use the definitions for the covariant vector
$$
R_p \eqdef \fr12\D{F^2(g;R)}{R^p} \equiv P_p
$$
and the FMT
$$
g_{pq}(g;R) \eqdef \fr12\, \fr{\prtl^2F^2(g;R)}{\prtl R^p\prtl
R^q} =\fr{\prtl R_p(g;R)}{\prtl R^q}.
$$

Thus we get the {\it Finsleroid-relativistic space
 }
  \be
\cE_g^{SR}=\{\cV_g;\,F(g;R);\,g_{pq}(g;R);\,R\in\cV_g\} \ee
and the
{\it Finsleroid-relativistic co-space } \be
\hat\cE_g^{SR}=\{\hat\cV_g;\,H(g;P);\,g^{pq}(g;P);\,P\in\hat\cV_g\}.
\ee



Special calculations can be used to verify the equalities
\ses\\
\be \det(g_{pq}(g;R))=-[j(g;R)]^8 \ee
and \be
\sign(g_{pq})=\sign(g^{pq})=(+ - - -). \ee

The following assertion is valid:
 \it for the Finsleroid space $\cE_g^{SR}$ the Cartan torsion tensor
$$
C_{pqr} \eqdef \fr12\D{g_{pq}}{R^r}
$$
is of the special algebraic form
 \rm
\be
C_{pqr}=\fr1N\lf(h_{pq}C_r+h_{pr}C_q+h_{qr}C_p-\fr1{C_tC^t}C_pC_qC_r\rg),
\ee \it
where
\be C_tC^t=-\fr{N^2g^2}{4F^2} \ee
and $N=4$. \rm

Proof is gained by straightforward calculations
on the basis of the explicit form of components of the FMT
and the Cartan tensor (see more detail in  [9--11]). Inserting
(A.20)--(A.21) in the general expression for the curvature tensor
$$
S_{pqrs} = C_{tqr}\3Cpts-C_{tqs}\3Cptr
$$
yields the following simple result after rather simple straightforward calculations:
\be S_{pqrs}=S^*
(h_{pr}h_{qs}-h_{ps}h_{qr})/F^2 \ee
with the constant
 \be S^*=\fr14g^2.
\ee \rm \ses
The tensor \be h_{pr} \eqdef g_{pr} -l_pl_r
\ee
has been used, where  $ l_p=R_p/F(g;R)$ -- the unit vector components.


The FMF  (A.12) defines the {\it pseudo-Finsleroid}
 \be
  \cF^{Relativistic}_g ~:=\{R\in \cV_g:F(g;R)\le 1\}.
   \ee
The associated
{\it
 indicatrix $\cI_g $} defined by
 \be
 \cI_g :=\{R\in \cV_g:F(g;R)=1\}
  \ee
is the surface of the pseudo-Finsleroid.
With the given FHF (A.14), the body
 \be
\hat\cF^{Relativistic}_g := \{\hat R\in \hat \cV_g: H(g;\hat
R)\le1\} \ee is called the \it co-pseudo-Finsleroid\rm.
The respective figuratrix introduced according to
 \be
  \hat\cI_g := \{\hat R\in \hat \cV_g: H(g;\hat
R)=1\}
 \ee
is called the  \it co-indicatrix\rm.

\ses

From (A.22)--(A.24) it follows that   \it in case of the Finsleroid space
 $\cE_g^{SR}$ the indicatrix is a space of the constant negative curvature
 which value is equal to
\rm \be R_{\rm Ind}=
-\lf(1+\fr14g^2\rg)\le-1. \ee
\ses

The respective    {\it equation of  $\cF_g$-geodesics} is of the form
\be \fr{d^2R^p}{ds^2}+\3Cqpr(g;R)\fr{dR^q}{ds}\fr{dR^r}{ds}=0, \ee
where $s$ is the parameter of the arc-length defined in accordance with
the rule
 \be ds
\eqdef \sqrt{ g_{pq}(g;R)dR^pdR^q}. \ee

The use of the functions
\be
A(g;R)=R^0-\fr12g\mR,\qquad \hat A(g;P)=P_0+\fr12g\mP \ee
and
\be
L(g;R)=\mR-\fr12gR^0,\qquad \hat L(g;P)=\mP +\fr12gP_0
\ee
is often convenient in various calculations.

In the limit  $g\to 0$
the considered space degenerates to the ordinary pseudoeuclidean case:
$$
 B|_{_{g=0}}= -[(R^0)^2-{\bf R}^2],
\quad \hat B|_{_{g=0}}= -[(P_0)^2-{\bf P}^2],
$$
\ses
$$
j|_{_{g=0}}= \hat j|_{_{g=0}}= 1, \, F|_{_{g=0}}=
\sqrt{|(R^0)^2-{\bf R}^2|},
$$
\ses
$$
H|_{_{g=0}}= \sqrt{|(P_0)^2-{\bf
P}^2|},
 \quad
g_{pq}|_{_{g=0}}= e_{pq},
$$
\ses
$$
g^{pq}|_{_{g=0}}= e^{pq}, \quad C_{pqr}|_{_{g=0}}= 0, \quad R_{\rm
Ind}|_{_{g=0}}= -1.
 $$
 Since at $g=0$ the space $\cE_g^{SR}$
is pseudoeuclidean,
then
 $
\cI_{g=0}
 $
is the ordinary unit hyperboloid.


\ses\ses

\setcounter{sctn}{2} \setcounter{equation}{0}

{\nin\bf Appendix B.  Quasi-pseudoeuclidean transformation} \ses
\ses\ses

Let us introduce the nonlinear transformation
\be t^i=\si^i(g;R) \ee \ses
with the functions \ses \be \si^0=
\lf|\fr{R^0+g_-|{\bf R}|}{R^0+g_+|{\bf R}|}\rg|^{-G/4}
\lf(R^0-\fr12g\mR\rg), \qquad \si^a= h\lf|\fr{R^0+g_-|{\bf
R}|}{R^0+g_+|{\bf R}|}\rg|^{-G/4} R^a; \ee
$i,j,...=0,1,2,3 $ and
$a,b,...=1,2,3$.
With the help of the transformartion, we can go over to from the  variables
$\{R^p\}$ to the new variables $\{t^i\}$. The inverse transformation
 \be
  R^p=
\mu^p(g;t) \ee
involves the functions
\be
 \mu^0=
\lf|\fr{t^0-m}{t^0+m}\rg|^{G/4} (t^0+\fr12Gm), \qquad \mu^a= \fr1h
\lf|\fr{t^0-m}{t^0+m}\rg|^{G/4} t^a, \ee
\ses\\
so that \be t^i\equiv\si^i\lf(g; \mu(g;t)\rg), \qquad
R^p\equiv\mu^p\lf(g; \tau(g;R)\rg). \ee The notation
$$
m=\sqrt{|r_{ab}t^at^b|}\in[0,\iy)
$$
has been used; the constant $h$ is given by the formula (A.4).

Let us introduce the pseudoeuclidean metric function
 \be
S(t)\eqdef\sqrt{|e_{ij}t^it^j|} \equiv \sqrt{|(t^0)^2-m^2|} \ee
 $
(e_{ij}={\rm diag}(1,-1,-1,-1) $ -- the pseudoeuclidean metric tensor).
It can readily be verified that the insertion of the the functions
 (B.2) in (B.6)
yields the identity
\be F(g;R)=S(t) \ee
with the function $F(g;R)$ which is exactly the FMF
 (A.12). In this way, we call
 (B.1)--(B.2)
the  \it quasi-pseudoeuclidean transformation\rm.


\ses\ses

The functions (B.2) are obviously homogeneous of degree 1
with respect to the variable ~$R$:
 \be
\si^i(g;bR)=b\si^i(g;R), \qquad b>0,
 \ee
from which it ensues that the derivatives
 \be
 t_p^i(g;R)\eqdef\D{\si^i(g;R)}{R^p} \equiv
\si_p^i(g;R)
 \ee
obey the identity \be t_p^i(g;R)R^p=t^i. \ee Calculating the determinant gives merely
 \be \det(t_p^i)=j^4h^3. \ee

Similarly, \be \mu^p(g;bt)=b\mu^p(g;t), \qquad b>0, \ee
 \ses
 \be
\mu_i^p(g;t)\eqdef\D{\mu^p(g;t)}{t^i},
 \ee
  and
   \be
\mu_i^p(g;t)t^i=R^p. \ee

Next, let us now construct the tensor \be n^{ij}(g;t)\eqdef t_p^it_q^jg^{pq}. \ee
Straightforward rather lengthy calculations
result in the following simple representations
 \ses \be n^{ij}(g;t)=h^2e^{ij}-\fr14g^2l^il^j,
\qquad n_{ij}(g;t)=\fr1{h^2}e_{ij}+\fr14G^2l_il_j \ee
\ses\\
($e_{ij}e^{jm}=\de_i^m$ и $n_{ij}n^{jm}=\de_i^m$),
where
 \be
l^i\eqdef t^i/S(t), \qquad l_i\eqdef e_{ij}l^j \ee
\ses\\
are respective pseudoeuclidean unit vectors.
For them the equalities
$$
l^il_i=1, \quad n_{ij}l^j=l_i, \quad n^{ij}l_j=l^i, \quad
n_{ij}l^il^j=1, \quad n_{ij}t^it^j=S^2
$$
are valid.
The inversion of (B.15) can be written in the form
\ses\\
\be g_{pq}(g;R)=n_{ij}\lf(g;\si(g;R)\rg)t_p^i(g;R)t_q^j(g;R). \ee
\ses\\
We also obtain \be \det(n_{ij})=-h^6. \ee


\ses

We call the tensor $\{n\}$ with the components (B.16)
 {\it quasi-pseudoeuclidean metric tensor},
and the very space \be \cK_g := \{\cQ;\,S(t);\,n_{ij}(g;t);\,t\in
\cQ\} \ee \it quasi-pseudoeuclidean space.
 \rm The formulas (B.7) и (B.15)
show explicitly that  space defined is
quasi-pseudoeuclidean image of the Finsleroid-relativistic
space $\cE_g^{SR}$, such that   {\it when using
the quasi-pseudoeuclidean transformations the studied Finsleroid-relativistic space
 $\cE_g^{SR}$
transforms in the quasi-pseudoeuclidean space
 $ \cK_g=\si(\cF_g)$  differed essentially from the pseudoeuclidean space
 $\cE\equiv\cK_{g=0}$. }

Let us evaluate from the tensor (B.16) the associated Christoffel symbols
 \be
\3Nmin=n^{mk}N_{ikj}, \quad
N_{ikj}=\fr12(n_{ik,j}+n_{jk,i}-n_{ij,k}). \ee
We have subsequently
 \be
n_{ik,j}\eqdef\D{n_{ik}}{t^j}=\fr14g^2(H_{ij}L_k+H_{kj}L_i)/S,
\ee
\ses \be H_{mi}=e_{mi}-L_mL_i \equiv h^2(n_{mi}-L_mL_i), \ee \ses
\be L^iH_{ij}=0, \ee \ses \be N_{mjn}=\fr14G^2H_{mn}L_j/S, \qquad
\3Nmin =\fr14G^2H_{mn}L^i/S, \ee
and
 \be
\3Nimj(t)=\fr14G^2L^mH_{ij}/S. \ee
This entails the properties
$$ t^i\3Nimj=0, \qquad \3Nijj=0. $$



\bigskip

\def\bibit[#1]#2\par{\rm\noindent\parskip1pt
                     \parbox[t]{.05\textwidth}{\mbox{}\hfill[#1]}\hfill
                     \parbox[t]{.925\textwidth}{\baselineskip11pt#2}\par}

\nin{\bf References}
\bigskip

\bibit[1] H.~ Busemann: \it Canad. J. Math. \bf1 \rm(1949), 279.

\bibit[2] H.~ Rund: \it The Differential Geometry of Finsler spaces, \rm
Springer-Verlag, Berlin 1959.

\bibit[3] G.S.~ Asanov: \it Finsler Geometry, Relativity and Gauge Theories, \rm
D.~Reidel Publ. Comp., Dordrecht 1985.

\bibit[4] D.~Bao, S.S.~ Chern, and Z.~ Shen: \it An
Introduction to Riemann-Finsler Geometry,
\quad\rm Springer, N.\.Y., Berlin, 2000.

\bibit[5] G.S.~ Asanov: \it Rep. Math. Phys. \bf 45 \rm(2000), 155;
\bf 47 \rm(2001), 323.

\bibit[6] G.S.~ Asanov:
arXiv:hep-ph/0306023, 2003;
arXiv:math.MG/0402013, 2004;

\end{document}